\documentclass[aps,prl,amsmath,amssymb,amsfonts,floatfix,reprint]{revtex4-1}
\usepackage[pdftex]{graphicx}
\usepackage[squaren]{SIunits}%\setstretch{1.5}
\usepackage{color}
\usepackage{textcomp} 
\begin{document}

\title{Transverse spin Seebeck vs.\,Anomalous and Planar Nernst Effects\\ in Permalloy Thin Films}

%\title{Nernst vs.\,Transverse Spin Seebeck Effects in permalloy Thin Films}

\author{M. Schmid$^{1}$}
\author{S. Srichandan$^{1}$}
\author{D. Meier$^{2}$}
\author{T. Kuschel$^{2}$}
\author{J.-M. Schmalhorst$^{2}$}
\author{M. Vogel$^{1}$}
\author{G. Reiss$^{2}$}
\author{C. Strunk$^{1}$}
\author{C.H. Back$^{1}$}

\affiliation{$^{1}$Institute of Experimental and Applied Physics, University of Regensburg, D-93040, Germany}
\affiliation{$^{2}$Thin Films and Physics of Nanostructures, Department of Physics, Bielefeld University, D-33501 Germany}
\date{\today}

\begin{abstract}
Transverse magneto-thermoelectric effects are studied in permalloy thin films grown on MgO and GaAs substrates and compared to those grown on suspended SiN$_x$ membranes. The transverse voltage along platinum strips patterned on top of the permalloy films is measured vs.\,the external magnetic field as a function of angle and temperature gradient. After the identification of the contribution of the planar and anomalous Nernst effects, we find an upper limit for the transverse spin Seebeck effect, which is several orders of magnitude smaller than previously reported.
\end{abstract}

\maketitle

The field of spin caloritronics \cite{bauer} connects spin and thermoelectric transport phenomena including spin dependent thermopower \cite{slachter}, thermal spin transfer torque \cite{hatami}, spin \cite{liu} and anomalous Nernst effects \cite{pu,huang} and spin Seebeck tunneling \cite{jansen}. A novel effect called spin Seebeck effect (SSE) has recently attracted attention since it may be appealing to control pure spin currents using thermal gradients. In general the SSE has been observed in two distinct sample configurations, namely transverse (TSSE) and longitudinal (LSSE). In case of TSSE a spin current is generated perpendicular to a thermal gradient $\vec{\nabla}T$. In the standard TSSE setup $\vec{\nabla}T$ is applied in the plane of an in-plane magnetized ferromagnetic layer (FM). In case of LSSE a spin current is generated parallel to $\vec{\nabla}T$, which is typically applied out-of-plane to a FM/NM bilayer. In both cases transverse voltages arising from the inverse spin Hall effect (ISHE) are measured along normal metal (NM) strips/layers, with high spin-orbit interaction \cite{silsbee,kimura}. For ISHE detection of a spin-imbalance in e.g. Pt, the transverse electromotive force is given by $E_{\mathrm{ISHE}} = D_{\mathrm{ISHE}} J_{\mathrm{s}} \times \sigma$, where $D_{\mathrm{ISHE}}$ is the ISHE efficiency, $J_{\mathrm{s}}$ the spin current entering the NM and $\sigma$ the spin polarization vector of the FM. TSSE has been observed in metallic \cite{uchida,uchida6,bosu}, semiconducting \cite{jaworski1} and insulating \cite{uchida2} ferromagnetic films. SSEs have been observed at low as well as at room temperature \cite{meier,jarowski2}.

The first observation of TSSE was on permalloy (Py) films on sapphire substrates \cite{uchida}. The authors suggested that the origin of the effect was the difference of the chemical potentials of the spin-up and spin-down electrons at the Fermi level. This explanation had to be discarded because the signal was observable at distances much larger than the spin diffusion length of the FM. Next the effect was detected in the diluted magnetic semiconducting GaMnAs \cite{jaworski1}. Small scratches interrupting the FM film did not change the signal, which indicates an important role played by the phonons in the substrate. SSE was observed in the insulating ferrimagnets LaY$_2$Fe$_5$O$_{12}$ (LSSE \cite{uchida2}) and Y$_2$Fe$_5$O$_{12}$ (TSSE \cite{uchida7}) where $J_{\mathrm{s}}$ can only be transported by magnons. These observations gave rise to a recent explanation of the SSE in terms of phonon-magnon and phonon-electron drag effects mediated through the substrate \cite{adachi}.

Notwithstanding the rapid advancement in this field, key issues of the SSE are not well understood. For instance there is little agreement on the value of the spin-Hall angle of Pt \cite{kimura,ando} leading to an uncertainty in the evaluation of the spin Seebeck coefficient $S_{\mathrm{S}}$. Additionally, in metallic FMs such as Py, other magneto-thermoelectric effects may contribute to the measured transverse voltage. Huang et\,al.\,\cite{huang} performed measurements on Py films deposited on bulk silicon substrates. Their results suggest that the transverse voltage generated by the out-of-plane temperature gradient due to the anomalous Nernst effect (ANE) in the bulk substrate overpowers the TSSE on the ferromagnetic thin film by a few orders of magnitude. In an attempt to resolve this question and understand the role of the substrate, Avery et\,al.\ conducted experiments on $500$\,nm thick suspended SiN$_x$ membranes in the transverse geometry \cite{avery}. In their study they observe the planar Nernst effect  \cite{vu}, i.e., the off-diagonal element $S_{xy}$ of the anisotropic magneto-thermopower (AMTEP) tensor $\mathbf{S}$, which is the thermal equivalent to the tensor of the anisotropic magnetoresistance (AMR). No detectable contribution of the TSSE was found in that study. Hence, a systematic study of Py films on bulk substrates as well as on suspended thin membranes is required, to elucidate the role of the substrate, and to discriminate the TSSE from other magneto-thermoelectric effects.

In this study we report on transverse thermoelectric effects in Py thin films on bulk MgO and GaAs substrates as well as on $100$\,nm thick SiN$_x$ membranes. The various effects contributing to the transverse voltage can be distinguished by virtue of the angular dependence of the transverse voltage and can be understood in terms of the AMTEP and the ANE, while possible contributions from the TSSE are significantly smaller than previously reported for samples on bulk substrates.

\begin{figure}
	\centering
		\includegraphics[width=8cm]{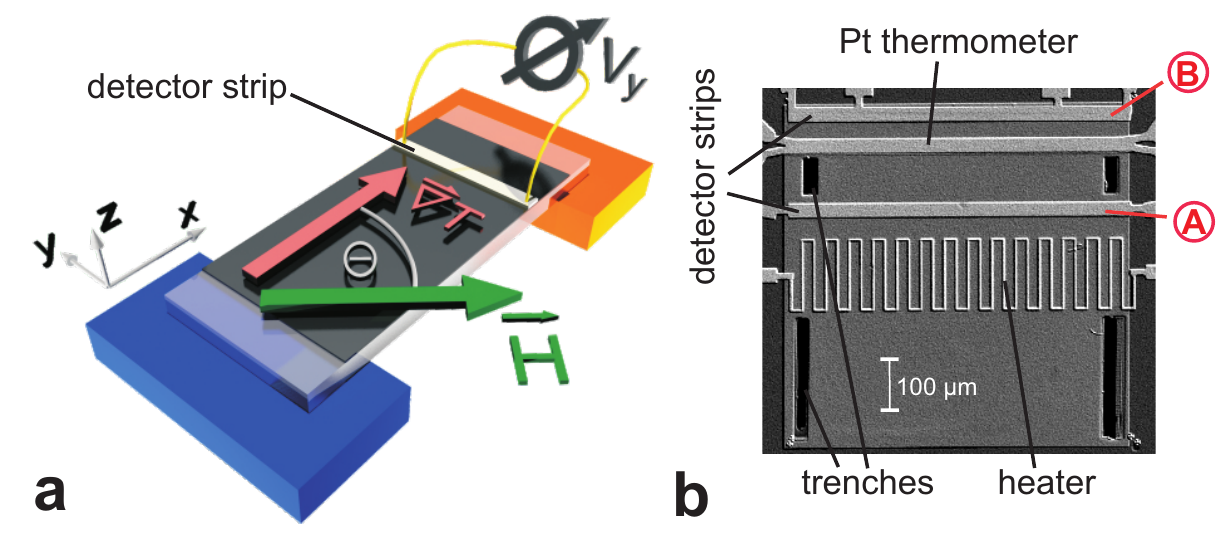}
		\caption{(a) Set-up~1: a Py film of 4\,$\times$\,8\,mm$^{2}$ size on a 4\,$\times$\,10\,mm$^{2}$ MgO substrate is subjected to a temperature gradient $\vec{\nabla}_{x}T$ along the $x$-axis between two Peltier elements with 7\,mm spacing. The transverse voltage $V_{y}$ is probed with bond wires on both ends of the Pt detector strip. The magnetic field $\vec{H}$ is applied in the film plane at an angle of $\Theta$ to the $x$-axis. (b) Set-up~2: a similar Py film patterned on a 500\,$\times$\,500\,\textmu m$^2$ SiN$_x$ membrane. Besides electrically isolated heater and thermometers two Pt detector strips labeled A and B are used.}
		\label{setup}
\end{figure}

Two types of set-ups were investigated for the experiments. Set-up~1: Py films of 20\,nm thickness were sputter deposited onto GaAs or electron-beam evaporated onto MgO substrates at a base pressure of $10^{-8}\,$mbar. Without breaking the vacuum, a single 100\,\textmu m wide and 10\,nm thick Pt strip was sputtered in situ through shadow masks onto the Py films, see Fig.\,\ref{setup}(a). In order to establish a controlled and reproducible temperature gradient, we mount the bulk substrate devices onto two temperature stabilized Peltier elements that are attached to a common copper block as heat sink allowing control of the temperature gradients for various base temperatures. The whole system is placed in a vacuum chamber with base pressure of $2\cdot10^{-6}$\,mbar in order to suppress any influence of convection. In control measurements the local temperatures on the film were determined independently using thermocouples. A pair of rotatable Helmholtz coils enables magnetic field sweeps in the sample plane at different angles $\Theta$ between the temperature gradient $\vec{\nabla}_{x}T$ and the magnetic field $\vec{H}$.  To minimize heat leaks by the electric contacts thin (50\,\textmu m) and $20$\,mm long Au bond wires were glued to both ends of the single Pt detector strip using silver epoxy spots with a diameter below 100\,\textmu m. The strip is located asymmetrically near one edge of the Py film.

Set-up~2: Bilayers with $20$\,nm of Py and $10$\,nm of Pt have been sputter deposited in-situ at a base pressure of $10^{-8}$\,mbar onto 500\,$\times$\,500\,\textmu m$^2$ large and $100$\,nm thick freely suspended SiN$_x$ membranes. Leaving two 20\,\textmu m wide Pt strips (A,B in Fig.\,\ref{setup}(b)) on the Py film, the rest of the Pt layer was removed by Ar etching. Contacts to these Pt detector strips have been made by a subsequent e-beam lithography step. A meandering $40$\,nm thick Au heater wire is located at the center of the membrane and close to one of the Pt strips. The heater and thermometers are electrically isolated from the Py film using $30$\,nm Al$_2$O$_3$. To provide a homogeneous temperature gradient between the heater and the heat sink trenches have been cut into the SiN$_x$-membrane (black regions) near the edges using a focused ion beam (see Fig.\,\ref{setup}(b)).
Set-up~2 was investigated in a different vacuum system at a base pressure of $10^{-6}$\,mbar.

Since for the observation of SSE the interfaces are important \cite{uchida2}, we verified the high quality of the FM/NM interfaces by performing spin pumping experiments on identical Pt/Py-bilayers and find a spin mixing conductance $g^{\uparrow\downarrow}\approx2\cdot10^{15}\,\mathrm{cm}^{-2}$.

\begin{figure*}
	\centering
		\includegraphics[width=18cm]{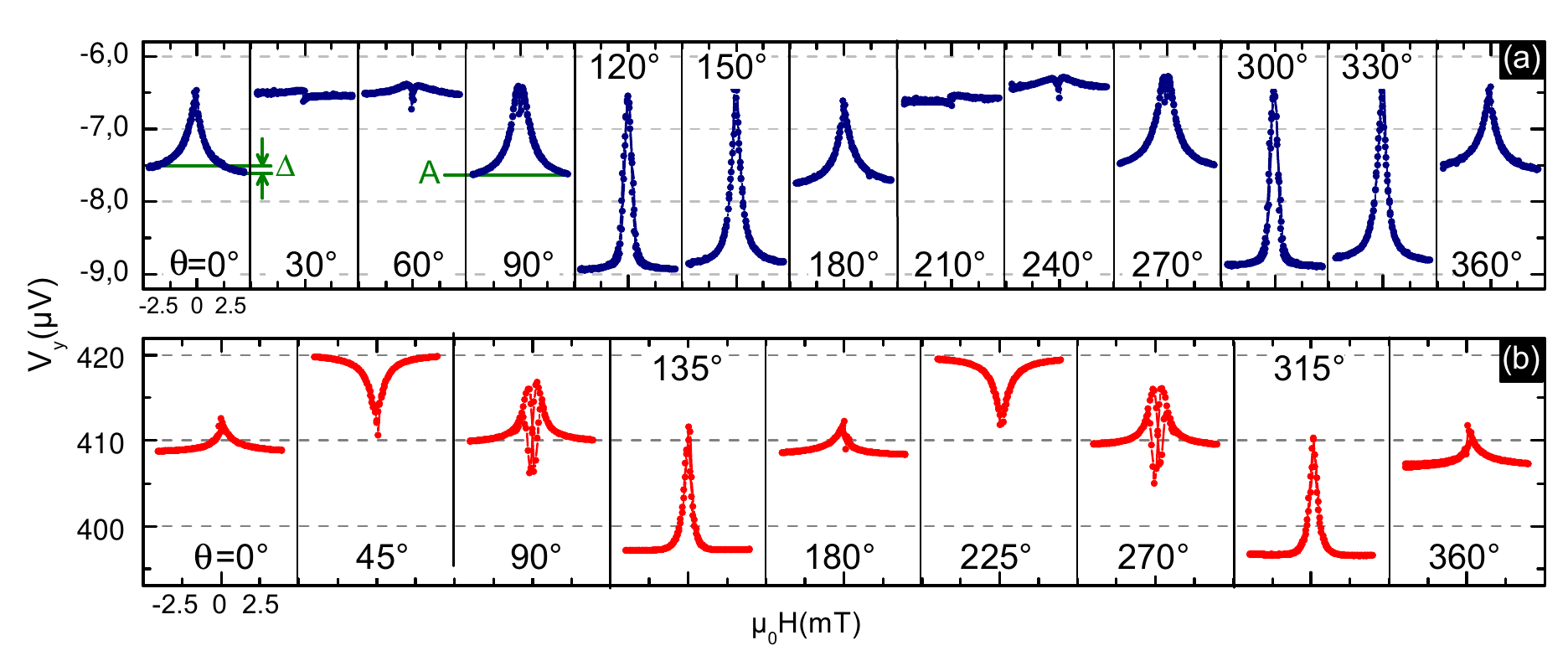}
		\caption{Transverse voltage $V_{y}$ vs.\ magnetic field and angle $\Theta$. (a) Set-up~1: Py on MgO substrate; $-4.5$\,mT $\leq$ $\mu_{0}H \leq$ $4.5$\,mT, $\vec{\nabla}_{x}T\approx$ $3.6$\,K/mm. The noise level in each sweep is 10 nV. (b) Set-up~2: Py on SiN$_{x}$-membrane; $-4$\,mT $\leq\,\mu_{0}H\ \leq4$\,mT, $\vec{\nabla}_{x}T\approx 240$\,K/mm is applied over a distance of 200\,\textmu m. Data is obtained at strip A in Fig.\,1(b). The noise level in each sweep is 50 nV.}
		\label{angle}
\end{figure*}

Fig.\,\ref{angle}(a) and (b) display the transverse voltage $V_{y}$ measured while sweeping the magnetic field at different angles $\Theta$ in set-up~1 and~2, respectively. We first focus on set-up~1 (see Fig.\,\ref{setup}(a)). The temperatures of the Peltier elements were set to $343$\,K (hot) and $293$\,K (cold). Taking account the thermal resistance of the junction interfaces this results in a temperature difference of (25$\pm$5)\,K ($\vec{\nabla}_x T\cong$3.6\,K/mm) across the Py film.  Fig.\,\ref{angle} shows three main contributions to $V_y(H)$: sharp maxima/minima located around $H=0$, an oscillation of the base line, i.e., the average $A(\Theta)\ =\ [V_y^{-}+V_y^{+}]/2$
of the saturation values $V_y^{\pm}$
evaluated for $H\gtrsim 4\;$mT, and a small asymmetry $\Delta(\Theta)\ =\ V_y^{-}-V_y^{+}$
 of $V_y^\pm$ with respect to the orientation of $\vec{H}$.

The prominent peak/dip structure near $H=0$ and the oscillation of the base line $A(\Theta)$ arise naturally from the off-diagonal element $S_{xy}(\Theta)$ of the tensor $\mathbf{S}$ of the AMTEP that locally connects the transverse electric field $E_y$ with $\vec{\nabla}_x T$.  $S_{xy}$ has to vanish, if $\vec{M}$ is aligned with one of the two principal axes of $\mathbf{S}$, i.e., when the angle $\phi$ between $\vec{M}$ and $\vec{\nabla}_x T$ is equal to $0^\circ, 90^\circ, 180^\circ$ or $270^\circ$ during the magnetization reversal process. Maximal (minimal) values of $V_y$ are observed, when $\phi=45^\circ, 225^\circ$ ($135^\circ, 315^\circ$). The variation of $V_y$ with $H$ at a fixed angle results from the rotation of $\phi$ during the magnetization reversal process. If $\Theta$ coincides with the orientation $\Theta_e$ of the easy axis of the film we have $\phi,\,\Theta=\,$const, and  $V_y(H)$ remains essentially constant. Hence, we can infer $\Theta_e\approx 35^\circ$ from Fig.\,\ref{angle}(a). The small uniaxial magnetic anisotropy arises from a residual magnetic field during deposition \cite{chikazumi,dillinger}.

\begin{figure}[b]
	\centering
		\includegraphics[width=7.5cm]{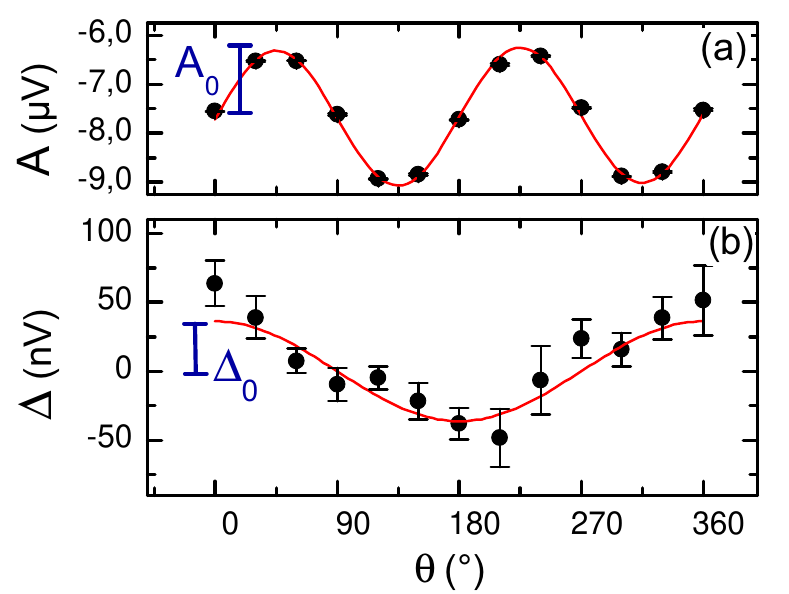}
	  \caption{Average A (a) and difference $\Delta$ (b) of the saturation values from Fig.\,\ref{angle}(a) as a function  of angle $\Theta$. The red lines show the $2A_{0}\sin(\Theta)\cos(\Theta)+c$ fit for (a) and the $\Delta_{0}\cos(\Theta)$ fit for (b).}
		\label{fits}
\end{figure}

As shown in Fig.\,\ref{fits}(a) the variation of the base line follows $A(\Theta)=2A_0\sin(\Theta)\cos(\Theta)+c$ \cite{vu}  with an amplitude $A_{0} \approx 1.2$\,\textmu V and a contribution $c$ that is independent of $\Theta$.
In contrast, the asymmetry $\Delta$=$\Delta_{0}\cos(\Theta)$ of the saturation values $V_y^\pm$ plotted in Fig.\,\ref{fits}(b) follows a simpler cosine dependence  with an amplitude $\Delta_{0} \approx 50$\,nV. The angle dependence of $\Delta$ is consistent with both the TSSE and the ANE.

\begin{figure}[b]
	\centering
		\includegraphics[width=8cm]{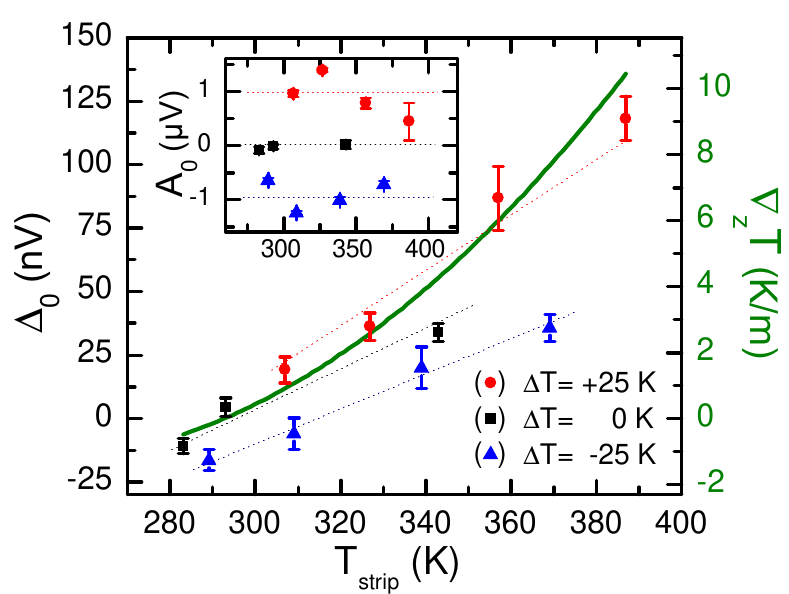}
		\caption{$\Delta_{0}$ vs.\ temperature at the Pt strip position for different $\vec{\nabla}_{x}T$. The inset shows the related average values $A_{0}$. The measurements are taken on a single MgO-substrate. The green line indicates the results of the finite element simulation. The dotted lines are a guide to the eye.}
		\label{mgo}
\end{figure}

Further measurements have been performed at different base temperatures with three fixed temperature differences $(\Delta T)_{x} =0$ and $\pm25$\,K. In Fig.\,\ref{mgo} we plot the resulting values of $\Delta_{0}$ and $A_{0}$ vs.~the local temperature at the Pt strip $T_{\mathrm{strip}}$. The inset shows values of $A_{0}$ with an average around $0.8$\,\textmu V for $(\Delta T)_{x} = 25$\,K with changing sign for opposite temperature gradient direction. Within the scatter the value of $A_{0}$ is essentially independent of base temperature. As expected for a thermo-voltage $A_0$ vanishes in the absence of an in-plane temperature gradient $[(\Delta T)_{x} = 0$\,K].

The asymmetry amplitude $\Delta_{0}$ in Fig.\,\ref{mgo} increases as a function of $T_{\mathrm{strip}}$. In contrast to $A_{0}$, $\Delta_{0}$ remains finite even for $(\Delta T)_{x} = 0$\,K, and is comparable in magnitude to the data for $(\Delta T)_{x} = \pm25$\,K. This proves that the monotonic increase of $\Delta_{0}$ with increasing $T_{\mathrm{strip}}$ cannot be attributed to the TSSE, but has to be related to yet another magneto-thermoelectric effect with the same $\cos(\Theta)$ symmetry.

To study the influence of the substrate on the presence of transverse voltages, additional measurements have been performed on GaAs with the same Py film dimensions as described above. The AMTEP exhibits qualitatively similar temperature dependencies as on the MgO substrate and also $\Delta_{0}$ shows a similar behavior as on MgO (see supplement).

As the origin of the finite $\Delta_{0}$($(\Delta T)_{x}=0$\,K), we suggest an out-of-plane temperature gradient $\vec{\nabla}_{z}T$ producing a transverse voltage $V_{y}$ due to the ANE. In order to estimate the values of $\vec{\nabla}_{z}T$ that may generate the measured $\cos\Theta$ dependence of the transversal voltage, a finite element simulation has been performed (see Fig.\,\ref{mgo} solid line and right scale bar). $\vec{\nabla}T_{z}$ of a few nK/nm turns out to be sufficient to generate voltages of the measured order of magnitude. The model includes surface to surface radiation of two black bodies, i.e., the sample surface and a heat sink at room temperature at a few cm distance simulating the geometry of set-ups. Taking the sample dimensions into account, $\vec{\nabla}_{z}T$ transforms to $V_{y}$ using $\vec{\nabla}V_{\mathrm{ANE}}=-\alpha\hat{m}\times\vec{\nabla}T_{z}$ with the magnetic unity vector $\hat{m}$ and the anomalous Nernst coefficient $\alpha=2.6$\,\textmu V/K \cite{huang}. The calculated values (Fig.\,\ref{mgo} solid line and left scale bar) are in good agreement with the experimental data concerning magnitude and temperature dependence.

In addition to the ANE contribution, the traces for $(\Delta T)_{x}= \pm25$\,K are shifted up (down) with respect to the $(\Delta T)_{x}=0$ K data (Fig.\,\ref{mgo}). This asymmetry of the order of 50 nV may be related to the TSSE which is expected to change sign for opposite $\vec{\nabla}_{x}T$.

To verify that this contribution indeed arises from the TSSE we exchanged the Pt strip with a Cu strip, as Cu should reduce the ISHE and thus the TSSE contribution drastically. The $\Delta_{0}$ signal again exhibits finite values for $(\Delta T)_{x}=0$\,K and the data points for opposite $(\Delta T)_{x}$ are still shifted by a few tens of nV probably due to the asymmetric positiong of the Pt strip (see supplement). Taking this into account we measure no TSSE contribution within our detection limit of $\Delta_{0}=20$\,nV. Taking this as an upper limit of TSSE we estimate the spin Seebeck coefficient $S_{\mathrm{S}}$ in the same way as Uchida et\,al.\,\cite{uchida}: $S_{\mathrm{S}}=2\Delta_{0}d_{\mathrm{Pt}}/(\Theta_{\mathrm{H}}\eta l_{\mathrm{Pt}}(\Delta T)_{x})$ with Pt thickness $d_{\mathrm{Pt}}$, the length of the Pt strip $l_{\mathrm{Pt}}$, the spin Hall angle $\Theta_{\mathrm{H}}=0.0037$, the spin injection efficiency $\eta=0.2$. We find an upper limit of $S_{\mathrm{S}}=-5.4$\,pV/K for our data, which is about two orders of magnitude smaller than what is reported for comparable temperature differences and sample dimensions \cite{uchida}.

To verify whether the observed effects depend on the phonons of a massive substrate, we deposited the Py film on a suspended $100$\,nm thin SiN$_{x}$-membrane (see Fig.\,\ref{setup}(b)).  For the suspended films all measurements have been performed at a base temperature of ($280$ $\pm$ $0.5$)\,K. Fig.\,\ref{angle}(b) shows the angular dependence of the voltage traces obtained from Pt strip (A) (close to the heater). Measurements with the Pt strip (B) (cold side) resulted in identical curves. The data obtained from the suspended film exhibit similar sinusoidal dependence as the Py film on bulk substrate (see Fig.\,\ref{angle}(a)) but with a magnetic easy axis $\Theta_e\approx190^{\circ}$. The data was treated as before: $V_y$ again follows $A=2A_{0}\sin(\Theta)\cos(\Theta)+ c$ with an amplitude $A_{0}$ of about $10$\,\textmu V (see supplementary material). The higher offset values result from different wirering conditions in set-up 2. In contrast to the samples on a bulk substrate, $\Delta$ shows no $\cos(\Theta)$ dependence. Within the estimated error $\Delta_{0}$ does not exceed $\approx 250$\,nV (see supplement). Using the value of $S_{\mathrm{S}} = -2$\,nV/K, given by Uchida et\,al.\,\cite{uchida} one would expect a TSSE contribution to $\Delta_{0}$ of around $2$\,\textmu V. This is clearly not observed. On the other hand starting with $\Delta_{0}=250$\,nV as experimental input, the resulting upper limit of $S_{\mathrm{S}}$ on the membrane is $-0.3$\,nV/K which is one order of magnitude smaller than published \cite{uchida}. In addition, we measured $V_y$ on the suspended Py films without the Pt strip and found very similar results (see supplement). This shows that a possible voltage resulting from TSSE is a minor contribution in our experiment.

The absence of TSSE in the measurements of Avery et\,al.\,\cite{avery} may be related to the much lower phonon density of states in the thin suspended SiN$_x$ membrane as it was proposed that the TSSE is a phonon mediated effect \cite{adachi}. In this study we confirm the absence of TSSE in suspended Py films on 5 times thinner SiN$_{x}$ membranes. In addition, the TSSE observed in our Py films on bulk MgO and GaAs substrates was at most a few tens of nV despite the fact that we ensured a high transparency of the Py/Pt interface. Such small values for Py films both on suspended SiN$_x$ membranes and on bulk substrates, suggest that the phonons in the bulk substrate are of minor relevance for the reported observations of the TSSE in Py films \cite{uchida,uchida6,bosu}, and that our observations can be attributed to a large extent to the anisotropic spin scattering in the Py.

In conclusion, we have shown that for Py/Pt films on both bulk substrates as well as membranes, the dominant contribution to the transverse signal obtained in a TSSE geometry results from the anisotropic magneto-thermopower. In addition, an appreciable contribution from the ANE originating from an out-of-plane temperature gradient is present. Independent of the presence or absence of phonons a contribution of the TSSE can be neglected within our detection limits.

The authors gratefully acknowledge financial
support by the DFG within SpinCaT (SPP 1538)
and the BMBF.

\appendix

\end{document}